\documentclass[number,citesort,dvips]{arxstspdf}
\usepackage{sgmlsup}
\usepackage{graphics}
\usepackage{flushend,stfloats}
\volume{24}
\issue{4}
\pubyear{2009}
\firstpage{530}
\lastpage{546}
\doi{10.1214/09-STS304}

\begin{document}
\begin{frontmatter}

\title{Using GWAS Data to Identify Copy Number Variants Contributing
to Common Complex Diseases}
\runtitle{Copy Number Variants}

\begin{aug}
\author[a]{\fnms{Sebastian} \snm{Z\"{o}llner}\ead[label=e1]{szoellne@umich.edu}\corref{}}
\and
\author[b]{\fnms{Tanya M.} \snm{Teslovich}}
\runauthor{S. Z\"{o}llner and T. M. Teslovich}

\affiliation{University of Michigan and University of Michigan}
\address[a]{Sebastian Z\"{o}llner~is~Professor, Department
of Biostatistics, Department of Psychiatry and Center for
Statistical Genetics, University of Michigan, 1420 Washington Heights, Ann
Arbor, Michigan 48109-2029, USA \printead{e1}.}
\address[b]{Tanya M. Teslovich is Research Fellow, Department of
Biostatistics and Center for Statistical Genetics, University
of Michigan, Ann Arbor, Michigan, USA.}

\end{aug}

%
\begin{abstract}
Copy number variants (CNVs) account for more polymorphic base pairs in
the human genome than do single nucleotide polymorphisms~(SNPs). CNVs
encompass genes as well as noncoding DNA, making these polymorphisms
good candidates for functional variation. Consequently, most modern
genome-wide association studies test CNVs along with SNPs, after
inferring copy number status from the data generated by high-throughput
genotyping platforms.

Here we give an overview of CNV genomics in humans, highlighting
patterns that inform methods for identifying CNVs. We describe how
genotyping signals are used to identify CNVs and provide an overview of
existing statistical models and methods used to infer location and
carrier status from such data, especially the most commonly used methods
exploring hybridization intensity. We compare the power of such methods
with the alternative method of using tag SNPs to identify CNV carriers.
As such methods are only powerful when applied to common CNVs, we
describe two alternative approaches that can be informative for
identifying rare CNVs contributing to disease risk. We focus
particularly on methods identifying \textit{de novo} CNVs and show that
such methods can be more powerful than case-control designs. Finally we
present some recommendations for identifying CNVs contributing to common
complex disorders.
\end{abstract}

%

\end{frontmatter}

\section*{Background}

Genome-wide association studies (GWAS) have successfully identified many
loci contributing to common complex diseases, and additional variants
continue to be identified as sample sizes increase. However, nearly all
common single nucleotide polymorphisms (SNPs) associated with complex
diseases have small effect sizes and explain only a small fraction of
the heritability of disease \cite{1}. Hence, it is prudent to consider other
types of heritable variation that may account for this unexplained
heritability. One promising candidate is copy number variation (CNV).

CNVs are segments of the genome that exist in different copy numbers in
the population. Traditionally, CNVs are defined to be at least 1 kb long
\cite{2}, but as the ability to detect these polymorphisms improves, shorter
segments are also considered. About 90\% of CNVs have two allelic states
\cite{3}. By comparison to the NCBI human reference sequence or to a
study-specific reference sample, such biallelic CNVs are classified as
deletions if the alternate allele carries fewer copies of the variable
sequence than the reference, and insertions (or duplications) when the
alternate allele contains more copies than the reference. The remaining
10\% of loci have copy number states not compatible with a two allelic
system, many of which may be explained by multiple overlapping CNVs
\cite{3}.

Some publications refer to CNVs with appreciable minor allele frequency
as copy number polymorphisms (CNPs), and genomic regions containing
multiple overlapping CNVs are called CNV regions (\mbox{CNVRs}). Here, we will
use the term CNV for all copy number polymorphisms. The cancer community
has introduced the term copy number alteration (CNA) for somatic copy
number variation; in the following we focus on germline CNVs.

CNVs are distributed ubiquitously throughout the genome, with a 25-fold
enrichment near segmental duplications \cite{4,5}. The reported
proportion of
the human genome covered by CNVs varies between 16\% \cite{5} and 5\%
\cite{3}.
Such discrepancies arise because most CNVs are rare. About 40\% of the
covered region described by Itsara et al. \cite{5} shows divergent copy
number in only one out of $\sim$2000 individuals; CNVs with minor
allele frequency (MAF) $>1\%$ cover less than 1\% of the human genome.
Therefore, the number of detected CNVs will depend strongly on the
sample size of the study; larger samples are likely to detect much
larger numbers of CNVs. Moreover, CNV allele frequencies correlate with
CNV location; CNVs near segmental duplications have higher average
population frequency than do CNVs at random loci in the genome \cite{5}.
Taken together, these results suggest that more genetic variation is
attributable to CNVs than to SNPs \cite{6}. While several studies have shown
that CNVs encompass genes less often than would be expected by chance
\cite{5,7}, up to $\sim$2900 genes overlap known CNVs \cite{2}. Several CNVs
have been shown to be associated with common disorders (reviewed below),
but generally, carriers of genes with aberrant copy number do not show
noticeable clinical phenotypes. The phenotypic impact of CNVs near or
within genes is generally unclear.

It is of great interest to understand the contribution of copy number
variation to phenotypic diversity in humans, and especially to the risk
of common complex disorders. Several specialized methods, such as BAC
Array Comparative Genomic Hybridization (CGH) \cite{8}, Representational
Oligonucleotide Microarray Analysis (ROMA) \cite{9} and Agilent CGH
\cite{10} have
been developed to detect CNVs. It is also possible to infer CNVs using
data from genome-wide genotyping arrays. Such approaches are inexpensive
and convenient, since vast amounts of data generated during GWAS are
already available for analysis. However, the optimal strategy for
evaluating such data is still an open question.

Below, we will explore existing methods and data that may inform such
strategies. After a brief characterization of genomic patterns of copy
number variation and reported associations between CNVs and common
disorders, we will discuss the signals generated by genotyping arrays
that can be used to identify CNVs, the methods that exploit one or more
of these signals, and possible pitfalls of these methods. Based on the
genomic patterns of CNVs and the performance of CNV detection methods,
we will discuss several strategies to identify CNVs contributing to
disease risk, and provide approximate power calculations. Throughout the
paper, we will focus on challenges of analyzing genotype data and
hybridization data such as generated from modern genotyping platforms.

\section*{Genomics of CNVs}

In the following, we provide an overview of the genomic characteristics
of CNVs cataloged thus far. To illustrate several of the described
patterns, we summarize data deposited in the Database of Genomic
Variants (DGV) \cite{11}, which describes $>$20,000 structural variants
identified in more than thirty independent studies. However, some of the
reported data sets may be conflicting, as many early studies had high
false positive and/or false negative rates, as well as limited ability
to accurately determine the boundaries of CNVs. As technology improves,
patterns are becoming more reliable.

Studies consistently report that CNVs are distributed ubiquitously
throughout the genome \cite{2,7,12,13} while being 25-fold enriched in
regions of segmental duplication \cite{5}. Approximately two-thirds of CNVs
in the DGV are deletions, and most studies included in the DGV report
more deletions than duplications. It is not clear whether this
difference reflects an actual excess of deletion polymorphisms, or
whether detection methods have more power to identify deletions. Such a
bias is plausible, as most CNV detection methods rely on hybridization
intensities, and the relative difference in intensity due to a deletion
is larger than that corresponding to a duplication. However, among CNVs~$>100$ kb in length, duplications are more frequent than deletions \cite{5}.

\begin{figure}

\includegraphics{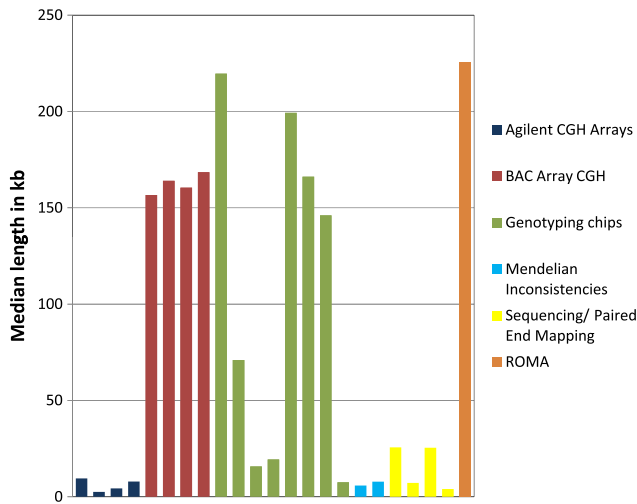}

\caption{ Median CNV length for 23 studies in the Database of Genomic
Variants. After excluding polymorphisms $<$1 kb in length, we selected all
studies with at least 50 polymorphisms remaining, and median CNV length
for each study is represented by one bar. Bars are color-coded to
indicate the method used to identify CNVs, as indicated by the label on
the right.}\label{fig1}
\end{figure}

The DGV contains CNVs as large as 8 Mb, with a median size of 17.6 kb.
The inferred length of a detected CNV is dependent on aspects of the
underlying technology, such as probe spacing, probe length and signal
resolution. To illustrate the differences between technologies, we
calculated the median CNV length for all studies collected in the DGV
(Figure~\ref{fig1}), excluding variants shorter than 1 kb. The median
length of
detected CNVs varies a great deal across studies, and the distribution
of CNV length suggests that BAC arrays and ROMA tend to overestimate CNV
size. Among studies that report at least 50 CNVs, the longest observed
median CNV length is 225 kb \cite{12}, while the shortest observed
median CNV
length is 2.5 kb \cite{14}. The median length of a study is clearly dependent
on its method of CNV detection. Agilent CGH methods, sequencing and
methods based on Mendelian inconsistencies estimate a median length of
$\sim$10 kb, while methods based on BAC CGH and ROMA \mbox{suggest} a median
length of $\sim$175 kb. Interestingly, methods based on SNP chips
generate widely varying estimates, ranging from 7.5 kb to 200 kb. Some
of this variability seems to be explained by differences between the
genotyping platforms and the resolution of the algorithms used to
analyze the data. As more recent experimental methods yield much shorter
estimates of median CNV length (even though they should be well powered
to detect longer CNVs), it seems likely that the CNV lengths reported
from BAC CGH arrays and some genotyping arrays are overestimates \cite{15}.

\subsection*{Origin of CNVs}

While CNVs are ubiquitous throughout the genome, we have only limited
understanding of their mutation process. The high frequency of CNVs in
regions of segmental duplication suggests that these CNVs are generated
by nonallelic homologous recombination \cite{16}. By careful analysis
of the
flanking sequence of 98 insertions and 129 deletions, Kidd et al. \cite{13}
determined that about 40\% of those CNVs were caused by nonallelic
homologous recombination. Of the remaining insertions, $\sim$30\% were
caused by nonhomologous end joining, $\sim$20\% by retrotransposition
and $\sim$10\% by expansion or contraction of a variable number of
tandem repeats. Among deletions, $\sim$45\% were caused by
nonhomologous end joining and $\sim$15\% by retrotransposition, while a
variable number of tandem repeat regions did not contribute. These
distributions depended on the size of the CNV; the proportion of CNVs
formed by nonallelic homologous recombination is larger among CNVs $>5$ kb.
In a recent study, Arlt et al. \cite{17} subjected human fibroblasts to
mitotic replication stress, which resulted in numerous copy number
changes. The authors observed that most breakpoint junctions showed
micro-homologies, suggesting that the copy number changes were generated
by nonhomologous end joining. It is not yet clear if the same processes
generate naturally occurring CNVs. Further work is necessary to estimate
the rates of these events and to understand the contribution of
surrounding genetic motifs. Such understanding may allow us to predict
mutation hotspots for CNV and to estimate mutation rates at these
locations. Based on these parameters, we can design methods to infer the
location of CNVs and hotspots of \textit{de novo} mutations. In fact,
several studies have used features of genomic DNA such as segmental
duplications to predict the locations of CNVs \cite{18,19}.

Current estimates of the rates of \textit{de novo} CNV mutations derive
from family studies. CNV status is inferred for members of a nuclear
family, and CNVs observed in the offspring, but not in the parents, are
assumed to be \textit{de novo} variants. As both false positives in the
offspring and false negatives in the parents result in false inference
of a \textit{de novo} event, it seems likely that the high rates of
\textit{de novo} events reported in some publications may be the result
of cell line artifacts or affected by the high error rates of the
applied CNV detection methods. In a recent study that carefully
controlled for such errors, McCarroll et al. \cite{3} observed 10
\textit{de
novo} events in 60 families, suggesting \textit{de novo} mutation rates
of $\sim$0.08 per generation per genome. When assessing \textit{de novo}
CNV mutation directly by sperm typing, Turner et al. \cite{20} estimated
rates between $5\bolds{\cdot}10^{-5}$ and $9\bolds{\cdot}10^{-7}$ per
genome per generation at four likely CNV mutation hotspots selected for
their high rates of nonallelic homologous recombination.

Rates of \textit{de novo} CNV mutation also are reflected in the extent
of linkage disequilibrium (LD) between CNVs and flanking markers. If a
CNV arises once during evolution, the LD pattern observed between the
CNV and nearby SNPs is expected to resemble the pattern of LD observed
between pairs of SNPs. On the other hand, if multiple mutational events
generate apparently identical CNVs, and each mutation event occurs on a
different haplotype background, we expect to observe little or no LD
between the CNV and adjacent markers. Several studies have suggested
that the extent of LD between CNVs and markers is comparable to the LD
between pairs of SNPs \cite{3,21}, implying a low \textit{de novo} mutation
rate of CNVs. CNVs in segmental duplications are reported to have less
LD with nearby SNPs \cite{22}. It is unclear whether this reduced LD is truly
caused by a higher rate of CNV mutation in these regions, or whether
this observation is an artifact of reduced SNP coverage. As SNP density
in segmental duplications is generally lower than in other genomic
regions due to the difficulty of designing high-quality genotyping
assays for duplicated SNPs \cite{22}, marker panels are less likely to
contain markers with the allele frequency necessary to obtain high
values of $r^{2}$. Nevertheless, coalescent simulations show that even
relatively high mutation rates of $10^{-5}$ are consistent with high
levels of linkage disequilibrium \cite{23}.

\subsection*{Frequency Spectrum and Signals of Selection}

Mutation rates of some CNVs are several orders of magnitude higher than
mutation rates of SNPs \cite{24}; therefore, it is remarkable that CNVs show
an excess of rare variants, compared to population genetic predictions
\cite{7}. Recently, Itsara et al. \cite{5} reported that in a sample of 2500
individuals, 35\% of all copy number variable sequence was copy number
variable in a single individual. Less than 1\% of CNVs had MAF~$>1\%$.
McCarroll et al. \cite{3} reported after analyzing the HapMap sample that
only 38\% of detected CNVs had MAF $>1\%$. The same paper \cite{3} emphasizes
that 8\% of CNVs responsible for interindividual variability have MAF~$\leq1\%$.
This estimate is again consistent with an excess of rare
variants; population genetics models of constant population size predict
that 2\% of mean difference between individuals will be generated by
polymorphisms with MAF~$\leq1\%$.

Nevertheless, CNVs with appreciable MAF occur worldwide. Jakobsson et
al. \cite{25} explored the distribution of 396 nonsingleton CNV loci inferred
in a worldwide sample of 405 individuals from 29 populations, observing
that 69\% of the detected CNVs occurred in more than one continental
group. Using the CNVs to form a population history, they recaptured the
same evolutionary history that is inferred from SNP data. In comparison,
Kidd et al. \cite{13} reported that of 1695 CNVs detected in a panel of four
Yoruba, two CEPH, one Chinese and one Japanese individual, 15\% of all
CNVs were observed in two or more continental groups. When analyzing the
HapMap sample using the Affymetrix 6.0 chip, McCarroll et al. \cite{3} found
that 42\% of all nonsingleton CNVs were present in more than one
continental group. While the differences between these estimates may be
a result of the different experimental platforms used, the common
message is that a large proportion of common CNVs can be found
worldwide. Whether this wide dispersal of common CNVs is the result of
parallel mutation in multiple ethnic groups or migration is not clear.

The frequency distribution of CNVs, with its strong excess of rare
variants, can be interpreted as a signal of purifying selection acting
on CNV loci, or as a signal of population growth. Under a model of
population growth we would observe similar allele frequency
distributions for CNVs and SNPs, as both are subject to the same
history. However, we observe a greater excess of rare variants among
CNVs than among SNPs, indicating that purifying selection is acting to
remove many derived CNV alleles from the population \cite{7,12}. This theory
is further supported by the finding that rare CNVs are more likely to
overlap with genes than common CNVs \cite{5}. Similar evidence has been
observed in model organisms: In inbred mouse strains, Henrichsen et al.
\cite{26} reported a paucity of CNVs in ubiquitously expressed household
genes and an excess of CNVs in genes with highly variable or
tissue-specific expression patterns as evidence that CNVs are under
purifying selection. Moreover, Emerson et al. \cite{27} reported evidence
that standing copy number variation in Drosophila is reduced due to
purifying selection. As the selection acting on CNVs is more pronounced
than that observed for SNPs, CNVs are likely to have greater functional
impact than SNPs, negatively affecting the reproductive fitness of
carriers.

\subsection*{Functional Signals of CNVs}

Given these signals for purifying selection, it is unsurprising that
several CNVs affecting the risk of common complex disorders have been
reported. Widely cited is the effect on HIV/AIDS risk of a copy number
polymorphism encompassing the gene encoding CCL3L1 \cite{28}, a potent human
immunodeficiency virus-1 (HIV-1)---suppressive chemokine and ligand for
the HIV co-receptor CCR5. Lower copy number of the CCL3L1 gene results
in reduced secretion of the CCL3L1 protein and is associated with
increased risk of HIV-1 infection. More recently, reduced copy number of
the beta-defensin gene cluster has been reported to be associated with
susceptibility to infectious and inflammatory diseases, particularly
Crohn's disease \cite{29,30} and psoriasis \cite{31}. Furthermore,
results of
Willer et al. \cite{32} implicated a 45 kb deletion upstream of NEGR1 as
being associated with body-mass index.

As most CNVs are rare, it can be difficult to demonstrate a
statistically significant association between a specific allele and
disease. Hence, some studies have examined the association between
disease status and total CNV load. Rather than testing for association
between a single CNV and a disease phenotype, such analyses assess
whether cases have a significant excess of CNVs (either deletions or
insertions) compared to controls. Using this design, Sebat et al. \cite{33}
demonstrated a contribution of deletions to the risk of autism. More
recently, Walsh et al. \cite{34} reported that \textit{de novo} insertions
contribute to the risk of schizophrenia, and Zhang et al. \cite{35} presented
similar results for bipolar disorder.

While several risk-CNVs have been detected, the mechanisms by which
these CNVs increase disease risk are largely unknown. Bridging the gap
between statistical association and biological understanding is
complicated by the fact that even CNVs that duplicate or delete entire
genes may not result in discernable phenotypes. Moreover, which genes
are affected by a CNV may be hard to predict. Recent studies comparing
gene expression and CNVs across twelve inbred strains of mice
demonstrated that, other than in the CCL3L1 gene, changes in copy number
often have little or no effect on expression levels \cite{26}. On the other
hand, the same study showed that longer CNVs can alter the expression of
genes over a distance of up to 3 Mb. Thus, CNVs that contribute to
disease risk may do so by acting on causal genes not normally associated
with the location of the CNV, creating yet another challenge as we seek
to understand the molecular mechanisms underlying disease risk.

A final challenge of detecting CNVs affecting common disorders is the
small effect size of such CNVs. Given prior genetic epidemiology
experiences with common complex diseases, we can make predictions about
possible effect sizes of CNVs under different scenarios. Consider a
common CNV (MAF $>5\%$) that is tagged by surrounding SNPs. If such a CNV
had a large effect size (OR $>2$), the surrounding SNPs would present a
strong signal for association in a GWAS, and the region would easily be
identified. So far, no such CNV has been detected; CNVs detected through
LD with neighboring SNPs have small effect sizes, comparable to those of
disease-associated SNPs \cite{30,32}.

Rare CNVs that segregate in the population are transmitted to offspring
according to Mendel's rules. Hence, CNVs with effect sizes comparable to
those of variants underlying Mendelian disorders are expected to
generate strong linkage signals. However, for the last 30 years,
geneticists have collected families for common complex disorders and
failed to identify linkage signals that can be explained by CNV. The
absence of strong linkage and association signals indicates that there
is an upper bound on the effect size of inherited CNVs that contribute
to complex traits. Population samples are unlikely to discover inherited
CNVs with large effect sizes.

\section*{Identifying CNVs in GWAS}

Given their genomic patterns, CNVs are enticing candidates for causative
variants, and it is of great interest to identify CNVs associated with
common diseases. As many CNVs are rare, and the effect sizes of common
CNVs are likely to be small, such studies require large sample sizes.
Genome-wide association studies that type densely spaced panels of SNPs
in large samples of cases and controls are already commonplace, and
therefore provide an inexpensive resource to explore the contribution of
CNVs to common diseases.

The utility of this approach depends on how many CNVs are covered by the
probes on genotyping arrays. Older genotyping arrays type relatively few
SNPs within common CNVs. As markers located within CNVs are likely to
fail multiple quality control criteria such as HWE, early array
designs excluded such ``problematic'' markers. Newer genotyping
technologies such as the Affymetrix 6.0 and the Illumina Human1M-Duo
BeadChip have increased coverage of CNV regions. Even most of the arrays
commonly used today directly interrogate only a subset of known CNVs.
McCarroll et al. \cite{3} reported that only 44\% of common CNVs
detected in
HapMap samples were represented by at least one SNP on the Affymetrix
500K or Illumina 650Y arrays, and less than 20\% of common CNVs are
represented by three or more SNPs. It has been estimated that at least
20\% of deletions longer than 1 kb span exactly zero probes on all
commercially available arrays \cite{36}. As accurate copy number estimates
require typing multiple SNPs within a CNV, the ability to infer CNVs
directly is limited by this coverage.

Most modern genotyping chips contain dedicated CNV probes to facilitate
copy number estimation. The Affymetrix 6.0 chip contains 800,000 probes
equally spaced over the genome, as well as 140,000 probes targeted
specifically at known CNV regions \cite{3}; the Illumina Human1M-Duo BeadChip
contains 36,000 nonpolymorphic probes to interrogate known CNV regions.
During analysis, such CNV probes can either be analyzed individually or
combined with SNP probes by treating CNV probes as genotyping probes
covering monomorphic SNPs. Independent of the specific platform, several
challenges must be overcome to perform thorough copy number analysis
using GWAS data. First, the signal is sparse; $>$99\% of each individual
genome is at normal copy number compared to a reference sequence.
Second, the signal is noisy, and a single SNP or probe is usually
insufficient to predict copy number status.

At least three types of evidence have been extracted from genotyping
data and used to infer the presence of CNVs: (1) Non-Mendelian
Inheritance errors (NMIs) in family data; (2) Departures from
Hardy--Weinberg Equilibrium (HWE); and (3) Differences in signal
intensity measured during the genotyping reaction.

\subsection*{Non-Mendelian Inheritance Errors (NMIs) in Family
Data}

\begin{figure}

\includegraphics{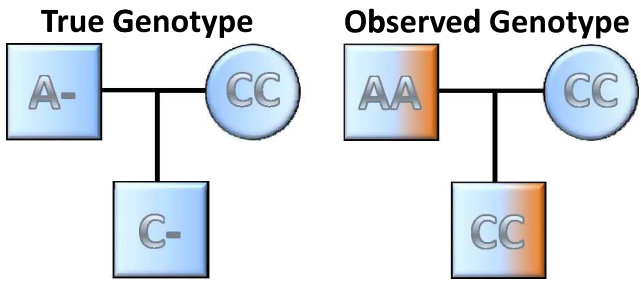}

\caption{Mendelian inheritance errors. The left panel displays the
genotype of a nuclear family at a single marker; the father is
hemizygous for a deletion that has been transmitted to the offspring.
The right panel shows the genotypes as they would be called by a
genotyping algorithm. Both hemizygotes are falsely typed as homozygotes.
Note that even though the actual transmission in the right panel follows
Mendel's rules, the observed genotypes seem to indicate an impossible
inheritance.}\label{fig2}
\end{figure}

Deletions segregating in families can cause the appearance of
non-Mendelian inheritance; hence, NMI analysis has proven to be a
powerful approach to localize deletions \cite{7,37}. In most genotyping
assays, hemizygous genotypes are inferred to be homozygous for the
present allele. If a hemizygous parent transmits the deletion-carrying
chromosome during meiosis, the child will be hemizygous and appear to be
homozygous for the allele transmitted from the other parent. If that
allele is different from the allele observed in the parent transmitting
the deletion, the offspring's genotype will be inconsistent with his two
parents under Mendel's rules, and the trio will be considered to be an
NMI (Figure \ref{fig2}). The observation of multiple consecutive SNPs with
non-Mendelian inheritance in the same trio indicates the presence of a
segregating deletion. However, not all deletions can be identified
through NMI analysis. Carriers in the parental generation will be
identified only if the chromosome carrying the deletion is transmitted
to the offspring. Even a transmitted deletion will generate an NMI only
if the allele transmitted from the other parent is inconsistent. The
probability of the deletion being transmitted is 0.5 and the probability
of a transmitted deletion creating an NMI is equal to the heterozygosity
of the SNP. Thus, the probability of observing an NMI if one of the
parents carries a deletion is equal to half the heterozygosity of the
SNP, therefore $\leq$0.25. Since consecutive SNPs covered by a deletion
are usually in LD, they do not generate NMIs independently of one
another even when conditioning on the deletion being transmitted. Hence,
the probability of seeing any pattern of NMIs among consecutive SNPs
depends on the haplotype frequencies in the population.

\subsection*{Departures from the Hardy--Weinberg Equilibrium}

Not only will genotyping algorithms generally identify hemizygous
individuals as homozygotes, they also will call homozygous SNPs in
duplicated regions as heterozygous, if different alleles are present at
the two different loci. Consequently, observed genotype frequencies of
SNPs covered by deletions or duplications may show departures from
HWE. SNPs within a deletion will show an excess of homozygous genotypes
for both alleles. Consecutive SNPs all showing an excess of homozygote
calls are indicators of a segregating deletion, and the minor allele
frequencies of SNPs within a deletion will be overestimated from the
data. The expected excess of homozygotes can be expressed dependent on
the frequency of the deletion, and the deletion frequency can be
estimated from the difference between the expected and the observed
number of homozygotes (see the \hyperref[app]{Appendix} for details).

For SNPs covered by duplications, the scenario is more complicated. SNPs
within a duplication usually will show an excess of heterozygous
genotype calls; the magnitude of this excess depends on the frequency of
the duplication, the distribution of alleles that have been duplicated
and the LD between the original region and the duplicate(s). Hence, the
frequency of a duplication cannot be estimated from genotyping data. In
the \hyperref[app]{Appendix}, we provide an overview of the change in genotype and
allele frequencies generated by this effect.

Note that such considerations assume uniform behavior of genotyping
algorithms. For some SNPs within a duplication, genotype clustering
algorithms may not be able to assign the correct three clusters to the
intensity signal, and produce false genotype calls or fail to call the
SNP. Thus, markers that fail quality control should be examined
carefully to determine whether they lie within CNVs.

\subsection*{Differences in Signal Intensity Measured During the
Genotyping Reaction}

Last, we can use the intermediate signal generated by modern genotyping
platforms to infer CNVs. The two most commonly used high-throughput
genotyping platforms (Illumina and Affymetrix) genotype by hybridizing
the DNA of an individual to a chip, generating a fluorescent signal for
each allele at every marker tested. The intensity of this fluorescent
signal depends on the number of alleles present. Due to the dynamic
range of modern arrays, which have been optimized to yield accurate
genotype calls, and since the scanners used to detect signal become
saturated, hybridization intensity is not quite proportional to the
number of copies of an allele. Moreover, the intensity distribution
varies between probes and between genotypes for each probe \cite{38}.
Consequently, it is not obvious how to model the distribution of
hybridization intensities. The intensity of the signal also depends on
all of the usual confounders of oligonucleotide array analysis such as
the total amount of DNA hybridized, background fluorescence and
hybridization quality \cite{39,40}. The signal distribution along a
chromosome has been described to show a wave-like pattern easily
mistaken for CNVs \cite{41}. Finally, interpreting this signal is challenging
because the inference of CNV status is confounded with the genotype
calling based on the same signal. Especially for low-quality DNA data
(e.g., from whole genome amplification), hybridization intensities are
often unsuitable to call CNVs although SNP genotype calls may be
accurate \cite{42}.

An important first step in the analysis of hybridization data is the
normalization of signal intensities. The raw data usually will consist
of one intensity signal for each of the two possible alleles. The goal
of the normalization step is to transform the two dimensional data into
a single random variable that is identically distributed for all loci
with baseline copy number, independent of the underlying genotype. For
Illumina \mbox{arrays}, this normalization is usually performed by calculating
the Log-R ratio. The calculation involves outlier removal, followed by
normalization against background signal. Based on these normalized
intensities, genotypes are called. The Log-R ratio (LRR) is the
logarithm of the ratio of the observed signal for a particular
individual to the average signal of individuals in the reference panel
with the same genotype. Hence, individuals with the same copy number as
those in the reference panel have LRR $\approx0$, while LRR $<0$
indicates a deletion, and LRR $>0$ indicates duplication. This
normalization algorithm assumes that individuals in the reference panel
have the baseline copy number for all markers. If this is not the case,
the normalization will be shifted, a problem, especially, for
individuals carrying rare alleles \cite{38}. In addition, Illumina's
normalization procedure provides the $B$ allele frequency (BAF), a measure
for the ratio of intensity signals between the two genotyping channels.
This statistic can be considered to be a quantitative representation of
genotype, taking values near 0 or 1 for homozygous genotypes and near
0.5 for heterozygous genotypes.

For Affymetrix arrays, no equivalent widely-used normalization strategy
exists. While quantile methods are most often used to normalize the
overall hybridization intensity across arrays (e.g., \cite{43}), most
methods analyzing Affymetrix data employ additional, method-specific
normalization algorithms to account for differences in hybridization
intensity distribution between loci and alleles.

\section*{Existing Methods for Analyzing GWAS~Data}

Two possible strategies exist for analyzing the contribution of CNVs to
common diseases in GWAS data. Either CNVs are tested using nearby SNPs
as proxies, or CNVs are inferred from genotyping data, and the resulting
calls are tested for association.

\subsection*{Linkage Disequilibrium (LD) Between CNVs and Nearby SNPs}

As common CNVs in unique regions of the genome are often in strong LD
with neighboring SNPs \cite{3,21}, these SNPs serve as proxies for the linked
CNVs, and SNP genotyping is an accurate and inexpensive alternative to
CNV typing. The utility of a SNP as a proxy measure is dependent on the
$r^{2}$ between the SNP and the CNV. For CNVs typed in the HapMap
sample~\cite{3} and other large population samples \cite{5}, it is possible to
define a
set of CNVs that are well tagged by known markers. McCarroll et al.
\cite{3}
reported that most common (MAF~$>5\%$), biallelic CNVs discovered in
HapMap samples can be captured perfectly by at least one SNP in the
HapMap Phase II data ($r^{2} = 1$ between CNV and tag SNP); however,
only 30--40\% of these CNVs are tagged perfectly by SNPs on commercially
available genotyping arrays, while 45--65\% can be captured by markers
with $r^{2} > 0.8$. Similarly, Cooper et al. \cite{36} reported that, among
84 common deletions observed in eight Yoruba, Japanese, Chinese and CEPH
samples (worldwide MAF~$>5\%$), 82\% were tagged by at least one HapMap
Phase II SNP with $r^{2} > 0.8$, and 48--54\% were captured ($r^{2} >
0.8)$ by markers on commercially available arrays. As standard operating
procedure, GWAS impute all HapMap markers, using algorithms such as MACH
\cite{44}, and test them for \mbox{association}. Consequently, GWAS studies already
test SNPs tagging most common CNVs. This strategy has been used
successfully to identify CNVs associated with complex traits \cite{30,32}.
However, this strategy has some weaknesses: First, tag SNPs cannot be
used to infer rare CNV alleles or \textit{de novo} events. Second, since
most markers on commercial arrays are biallelic, multiallelic CNVs are
necessarily poorly tagged. Third, CNVs located in segmental duplications
are generally more difficult to tag \cite{3,4}. For these reasons, the copy
number status of many CNVs must be estimated using other methods.

\subsection*{Analyzing CNV Calls}

Early approaches for identifying CNVs from genotyping data focused
largely on NMIs and departures from HWE to identify deletions in
HapMap samples \cite{7,37}. Kohler and Cutler \cite{45} combined NMIs,
deviations
from HWE and frequency of missing data to infer deletions from GWAS
data.

Presently, most CNV detection methods focus on analyzing hybridization
intensity data, often ignoring other sources of information such as LD
or departure from HWE. To identify CNVs, researchers adapted several
methods that were originally designed to analyze cancer data (e.g.,
circular binary segmentation, CBS \cite{46}) or designed for other platforms.
The first method specifically designed for genotyping arrays is an
extension of the SW-ARRAY algorithm \cite{47} by Komura et al. \cite
{48} to
analyze data from Affymetrix 500K chips. In the recent literature,
hidden Markov model (HMM) methods are the most commonly applied tool.
First proposed by Fridlyand et al. \cite{49}, these methods exploit the local
correlation of trait status. As CNVs often extend over multiple markers,
combining information across neighboring markers is often more powerful
than looking at one marker at a time. Colella et al. \cite{50} proposed an
objective Bayes Hidden Markov Model to infer location and carrier status
of CNVs from Illumina BeadArray data. With PennCNV \cite{15}, Wang et al.
extended this model to utilize information for related individuals. Such
HMM methods have been applied in many projects (i.e., \cite{5,25}), and most
CNVs in the databases have been located with these or similar
algorithms. Unfortunately, HMM methods have relatively high error rates,
especially for shorter CNVs. PennCNV has an error rate of 25\% for CNVs
of any length and 9\% for CNVs encompassing ten or more SNPs \cite{15}. While
no error rates have been reported for other HMM methods, they are not
fundamentally different from PennCNV and, it is unlikely that
they perform substantially better.

Most HMM and CBS methods used to infer CNVs analyze one individual at a
time, and only post-hoc combine the calls across individuals. While this
keeps the memory requirements for each analysis to a minimum, it
potentially reduces the ability to exploit occurrences of the same CNV
in multiple individuals. Recently, methods designed under a different
paradigm have been published. Rather than scanning the genome for
signals of copy number variation, these methods only analyze known
copy-number variable regions. Such methods do not have to account for
the \mbox{uncertainty} of the CNV location, and can therefore generate more
precise estimates of carrier status. The algorithm Canary \cite{43}
fits a
Gaussian Mixture model to the intensity distribution and assigns copy
number status according to cluster membership. Other recent methods
attempt to quantify the uncertainty of the CNV call; such measures of
uncertainty can be incorporated into tests for association by weighting
each call according to its confidence. CNVEM \cite{38} is based on a similar
idea, using a Bayesian framework to calculate the posterior probability
of copy number, thus accounting for the uncertainty in the CNV
genotyping. Similarly, Barnes et al. \cite{51} proposed a frequentist method
of modeling copy number states as a latent variable and then using a
mixture model to test for association. All of the methods focusing on
known CNVs report substantially lower error rates compared to HMM
models, although few such estimates of error rates have been replicated
independently.

Such methods for calling known CNVs depend on precise estimates of CNV
location. Large collections of CNVs have been described in multiple
databases, including the Database of Genomic Variants (DGV)~\cite{52}, the
Human Genome Structural Variation Project \cite{53} and the Copy Number
Variation Project Data Index~\cite{54}. Some care must be taken when
selecting loci from these databases; as technology and algorithms used
to detect CNV are still evolving, these databases contain false
positives, and not all common copy number variants have been detected
and reported. Furthermore, the boundaries of CNVs in these databases may
be imprecise, as some methods for CNV detection only yield approximate
boundaries. In practice, it may be advisable to focus on CNV collections
reported by recent studies, as these tend to be based on more precise
methodology. Of course, focusing on CNVs reported in databases is not
appropriate when exploring the impact of \textit{de novo} CNV mutations,
since such variants may not have been previously observed. In this case,
the analysis can be performed in two steps, with an initial CNV
discovery step using an HMM such as PennCNV. While such methods may not
detect every CNV in each carrier, it is sufficient to identify each CNV
once in the sample and to generate estimates of its borders. In a second
step, the copy number status of these CNVs can be called in all
individuals using more precise algorithms.

\subsection*{Comparing Tag SNPs and CNV Calling}

It is not obvious whether directly estimating CNV carrier status is
actually a useful strategy if a CNV is tagged by nearby SNPs; even in
the best case, methods estimating the carrier status of a CNV have much
higher error rates than SNP genotyping \cite{43}. The answer depends on the
degree of LD between the CNV and the proxy SNPs, as well as the error
rate for inferring CNVs directly. Here we assess which approach is more
powerful, by determining the sample size inflation necessary to overcome
power loss due to errors in CNV inferences, and compare it to the
inflation in sample size necessary to overcome the power loss due to
incomplete LD ($r^{2} < 1)$. We show that under many scenarios, testing
tag SNPs results in a more powerful test than calling CNVs and testing
inferred CNV calls.

Following an argument from Pritchard and Przeworski \cite{55}, we
derive the
distribution of a $\chi^{2}$-test for association based on a $2\times2$
contingency table dependent on the rate of calling error. Based on that
distribution, we calculate the inflation factor ($\mathit{IF}$) by which
the sample size needs to be increased to overcome the loss of power due
to CNV calling errors. Assuming no calling error, the distribution of a
$\chi^{2}$-test in a sample of $N_{1}$ cases and $N_{1}$ controls is
\[
\chi_{1}^{2} = \frac{( P(C|\mathit{case}) - P(C|\mathit{control}) )^{2}N_{1}}{2P(C)(1 -
P(C))},
\]
where $P(C|\mathit{case})$ is the observed frequency of the minor CNV allele in
cases, $P(C|\mathit{control})$ is the observed frequency of the minor CNV allele
in controls and $P(C)$ is the overall observed frequency of the minor
CNV allele. Now let $O$ be the event of calling the minor allele of the
CNV, $C$ the event of the minor CNV allele being present and $A$ the
event of the major CNV allele being present. Then we can parameterize
$P(O|C)$ as the probability of correctly calling the minor allele if the
minor allele is present and $P(O|A)$ as the probability of falsely
calling the minor allele if the major allele is present. Hence,
\[
P(O) = P(O|C)P(C) + P(O|A)P(A)
\]
is the total number of CNVs being called in the sample. In this model,
the $\chi^{2}$-test for association in a sample of $N_{2}$ cases and
$N_{2}$ controls is
\begin{eqnarray*}
\chi_{2}^{2} &=& \bigl( [P(O|C) - P(O|A)]\\
&&{}\hspace*{3pt}\times[P(C|\mathit{case}) - P(C|\mathit{control})]
\bigr)^{2}N_{2}\\
&&{}/\bigl({2P(O)\bigl(1 - P(O)\bigr)}\bigr).
\end{eqnarray*}
To calculate the increase in sample size necessary to overcome the loss
of power due to errors in calling CNV alleles, we can calculate the
inflation factor ($\mathit{IF}$):
\[
\mathit{IF} = \frac{N_{2}}{N_{1}} = \frac{1}{( P(O|C) - P(O|A)
)^{2}}\frac{P(O)(1 - P(O))}{P(C)(1 - P(C))}.
\]
The right side of the equation indicates the factor by which the sample
size has to be multiplied to overcome the loss of power due to calling
errors. This inflation factor can be directly compared to the inflation
of sample size necessary to overcome incomplete LD ($r^{2} < 1)$, as
testing for association at a marker with $r^{2} = x$ to the risk variant
inflates the sample size by~$1/x$ \cite{55}.

%
\begin{table*}[t]
\caption{Impact of calling error on association testing for
common CNVs. We display the inflation factor ($\mathit{IF}$) for sample
size, necessary to overcome typing error of common CNVs. The first line
shows the sample frequency of the rare CNV allele, the first column
shows the probability of falsely calling the rare CNV allele and the
second column shows the probability of correctly calling the rare CNV
allele. For each set of parameters the table shows the inflation factor
($\mathit{IF}$) for the sample size to overcome the effects of this genotyping
error and the LD ($r^{2}$) to a~tag SNP that results in the same loss of
power}
\label{tab1}
\begin{tabular*}{\textwidth}{@{\extracolsep{\fill}}lccccccccc@{}}
\hline
  &   & \multicolumn{8}{c@{}}{$\bolds{P(C)}$ \textbf{(freq. of minor CNV allele)}}\\
\ccline{3-10}\\[-6pt]
$\bolds{P(O|A)}$ & $\bolds{P(O|C)}$& \multicolumn{2}{c}{\textbf{0.02}} & \multicolumn{2}{c}{\textbf{0.05}} & \multicolumn
{2}{c}{\textbf{0.1}} & \multicolumn{2}{c@{}}{\textbf{0.2}}\\
\ccline{1-2,3-4,5-6,7-8,9-10}\\[-6pt]
\textbf{(false positive rate)}&\textbf{(sensitivity)} & $\bolds{\mathit{IF}}$ & $\bolds{r^{2}}$
& $\bolds{\mathit{IF}}$ & $\bolds{r^{2}}$ & $\bolds{\mathit{IF}}$ &
$\bolds{r^{2}}$ & $\bolds{\mathit{IF}}$ & $\bolds{r^{2}}$\\
\hline
0.01 & 0.9 & 1.74 & 0.57 & 1.37 & 0.73 & 1.25 & 0.80 & 1.20 & 0.83\\
& 0.8 & 2.05 & 0.49 & 1.59 & 0.63 & 1.44 & 0.69 & 1.40 & 0.71\\
& 0.7 & 2.49 & 0.40 & 1.88 & 0.53 & 1.70 & 0.59 & 1.66 & 0.60\\[5pt]
0.05 & 0.9 & 4.41 & 0.23 & 2.45 & 0.41 & 1.80 & 0.56 & 1.48 & 0.67\\
& 0.8 & 5.51 & 0.18 & 2.99 & 0.33 & 2.16 & 0.46 & 1.78 & 0.56\\
& 0.7 & 7.13 & 0.14 & 3.77 & 0.27 & 2.68 & 0.37 & 2.18 & 0.46\\
\hline
\end{tabular*}
\end{table*}

To compare tagging strategies with direct calling of CNVs, we calculated
the inflation factor for a range of error rates and CNV frequencies
commonly reported in the literature (Table \ref{tab1}). Most CNV calling methods
have reported error rates between 0.1 and 0.3. As falsely calling the
rare allele of a CNV at a specific location is unlikely under most
methods, we assumed that most errors were false calls of the major
allele in the presence of the minor allele; the probability of such
errors is ($1-P(O|C));$ we consider values of $P(O|C)$ between 0.7 and
0.9, for values of $P(O|A)$ of 0.01 and 0.05. For larger values of
$P(O|A)$, the inflation factor increases rapidly (data not shown). For
comparison, we also calculated the $r^{2}$ between the CNV and the best
tag SNP that results in the same inflation factor for the tag SNP
approach.

Our results indicate that calling error reduces the power of testing
rare CNVs more than it reduces the power of testing common CNVs. Even
modest error rates [$P(O|A)=0.01$, $P(O|C)=0.8$] increase the required
sample size for finding rare CNVs ($\mathrm{MAF} = 0.02$) by 50\% or more,
particularly relevant as large sample sizes are required to detect these
rare variants in the first place (Table \ref{tab1}). Comparison with LD
statistics indicates that, under these conditions, a SNP tagging the CNV
with $r^{2} \geq0.49$ is sufficient to provide a more powerful test
than inferring CNV status and directly testing the CNV for association
with disease. Furthermore, a high false positive rate [$P(O|A)$]
increases the sample size more than does a high false negative rate
($1-P(O|C)$). For high values of $P(O|A)$, inferring and testing a CNV
yields poor results, compared to the tagging method; under all
considered parameter combinations, a tag SNP with $r^{2} \geq0.67$ to
the CNV allows for a more powerful test for association.

Note that these considerations assume that only a single tag SNP
provides information about the allelic state of the CNV. In practice, we
can expect multiple SNPs to be in LD with the CNV, and combining
information across tag SNPs will result in an even more powerful test
statistic. However, when no tag SNP is available for a particular CNV,
valuable information may be gained by inferring CNV status directly from
GWAS data.

\section*{Testing CNVs for Association with Disease}

After inferring carrier status, several methods can be used to test for
association between inferred carrier status and disease. As most CNVs
are biallelic, we can apply methods developed for rejecting the null
hypothesis of no association between a biallelic marker and a phenotype,
such as the chi-square test or logistic regression. In such studies we
consider the inferred carrier status to be the true carrier status.
However, in tests for transmission distortion \cite{56,57} it should be
considered that transmitted CNVs are generally easier to detect than
nontransmitted CNVs, particularly if NMIs are used to identify
carriers. A further potential problem may be generated by the stringency
of the CNV calling algorithm. Commonly, such algorithms impose a high
burden of proof (e.g., posterior probability $>0.95$) before assigning the
minor allele, in order to minimize the effects of measurement error.
This approach can increase the number of false negative calls and
introduce nonrandom missingness, thus inflating the false-positive rate
of a family-based test for association \cite{51}.

Tests for association can be improved by accounting for the uncertainty
in the estimate of carrier status. Bayesian methods will generally
provide a posterior probability for carrier status \cite{38,51}, and in
frequentist inference methods this uncertainty can be ascertained by
bootstrap or jackknife procedures. Once this uncertainty is known, tests
for association can be adjusted accordingly. For Bayesian estimates we
can compare the summed expected posterior carrier status in a $\chi^{2}$
test or in a logistic expression.

Finally, Stranger et al. \cite{58} skip the step of inferring CNV
status for
such regions and directly test for association between hybridization
intensity and case-control status. This method is susceptible to false
positives due to shifts in mean and/or variance of the underlying
intensity distributions, and such shifts occur frequently in practice
\cite{51}.

\section*{Alternative Strategies to Association Mapping}

As most CNVs have low MAF, tests for association between a single CNV
and a disease are likely to have low power, especially if $p$-values are
corrected for multiple tests. Therefore, alternative strategies must be
considered. Here we present two such strategies: first a test for an
excess of \textit{de novo} CNV mutations at a locus and second a test
for an excess loading of minor CNV alleles in cases compared to
controls.

\subsection*{Detection of de novo CNVs}

As recent results indicate that \textit{de novo} CNV mutations are rare
\cite{3}, multiple \textit{de novo} mutations in the same region of the
genome suggest candidates for risk variants. However, even if \textit{de
novo} mutations are over-represented and highly penetrant among cases,
the combined variants at one locus are still unlikely to have allele
frequency $>1\%$ in cases. Hence, applying standard testing strategies to
compare allele frequencies between cases and controls will be
underpowered. Consider, for example, a genomic region carrying 6
\textit{de novo} mutations in 1000 cases, and none in 1000 controls.
Testing for association yields a Fisher's exact $p$-value of 0.015, no
clear evidence of association. However, this $p$-value does not account
for the observation that the rate of \textit{de novo} mutations is low
and therefore the probability of observing 6 \textit{de novo} deletions at
the same locus by chance is unlikely.

Nevertheless, it is not clear how many \textit{de novo} mutations must
be observed in the same region before the finding is significant. Such a
critical value depends on two parameters: the rate of \textit{de novo}
mutation, and the number of locations in the genome where such mutations
occur. While the mutation rate can be estimated from existing data sets,
early estimates of these rates were confounded with high false negative
rates \cite{59}. Recent studies suggest that these rates may be as low as
0.08 per genome per meiosis \cite{3}.

Estimating the second parameter, the number of genomic regions
experiencing \textit{de novo} CNV mutation, is more challenging. About
40\% of CNVs are generated by nonallelic homologous recombination \cite{13},
which is caused by flanking repetitive elements and segmental
duplications. Such segmental duplications cover 5\% of the genome. Thus,
it is likely that most nonpathogenic \textit{de novo} mutations occur in
only a small subset of the genome. However, better understanding of the
processes generating CNVs are necessary for a precise estimate of the
subset of the genome with high CNV mutation rate.

To explore the power of detecting a risk variant by observing an excess
of \textit{de novo} mutations, we performed computer simulations based
on the two parameters described above. We set the \textit{de novo}
mutation rate of noncausal CNVs to $\mu$ per meiosis per genome,
uniformly distributed over $k$ locations in the genome. We further
assumed that a subset $\varepsilon$, of all \textit{de novo} CNV alleles
would be identified. We did not model other sources of error, as false
positive \textit{de novo} CNV calls, by definition, are not expected to
cluster at particular loci in the genome, and therefore do not affect
our test statistic.

%
\begin{table*}[b]
\caption{Critical values and power in tests for de novo
mutations. The first row indicates the number of sites for
noncausal CNVs, and the second line displays the sample size. The third
line shows the number of CNVs in one location constituting a significant
number of de novo CNVs. The next three lines show the power to
detect a locus carrying an excess of de novo mutation, assuming
that these de novo mutations can be observed in 1\%, 0.5\% or 0.25\%
of all cases. The last column provides the sample size necessary to
detect the CNV with 80\% power in a case-control design at a
significance level of $10^{-5}$}\label{tab2}
\begin{tabular*}{\textwidth}{@{\extracolsep{\fill}}lccccccc@{}}
\hline
& \multicolumn{6}{c}{\textbf{Number of \textit{de novo} CNV mutation hotspots}}
& \\
\ccline{2-7}\\[-6pt]
& \multicolumn{3}{c}{\textbf{500}} & \multicolumn{3}{c}{\textbf{2000}} & \\
\ccline{2-4,5-7}\\[-6pt]
\textbf{Sample size} & \textbf{500} & \textbf{1000} & \textbf{2000} & \textbf{500} & \textbf{1000} &
\textbf{2000} & \textbf{Case-control sample size}\\
\hline
$CV_{\alpha}$ & 4\phantom{00.} & 5\phantom{00.} & 6\phantom{0} & 3\phantom{00.} & 4\phantom{00.} & 5\phantom{0} & \\
1.00\% & 0.52 & 0.86 & $>\!0.99$ & 0.72 & 0.94 & $>\!0.99$ & \phantom{0.}5500\\
0.50\% & 0.12 & 0.32 & \phantom{0.}0.76 & 0.29 & 0.52 & \phantom{0.}0.94 & 11,000\\
0.25\% & 0.02 & 0.04 & \phantom{0.}0.18 & 0.07 & 0.12 & \phantom{0.}0.52 & 22,000\\
\hline
\end{tabular*}
\end{table*}

Assuming a sample of $n$ nuclear families, we modeled as the null
distribution the total number of \mbox{detected} noncausal \textit{de novo}
CNVs, $c_{i}$, at each location~$i$, as Poisson-distributed with
rate $2\mu n\varepsilon/k$. Defining $M=\operatorname{max}\{c_{i}:i=1,\ldots,k\}$
as the maximum number of noncausal \textit{de novo} mutations observed
anywhere in the genome, the critical value for a test of an excess of
\textit{de novo} CNVs is equal to $CV_{\alpha} =\operatorname{min}\{x: P(M \geq x) <
\alpha\}$. Note that the alpha level chosen here is the experiment-wide
type I error rate; by maximizing $M$ over all CNV mutation hotpots, we
have corrected for genome-wide multiple testing.

To simulate the distribution of causal \textit{de novo} mutations at a
risk locus, we set $p$ as the proportion of cases in the population
carrying the \textit{de novo} CNV mutation at a specific risk locus.
Then the power for a test of excess \textit{de novo} mutations can be
calculated using the binomial distribution, $B(\varepsilon p,n)$.

We assessed the power of this method to detect a significant excess of
\textit{de novo} mutations. We first calculated critical values for
sample sizes of $n = 500$, 1000 and 2000 nuclear families, assuming a
\textit{de novo} mutation rate of noncausal CNVs of $\mu= 0.1$ per
meiosis per genome, uniformly distributed over $k = 500$ or 2000
locations in the genome. Assuming an average length of 50 kb per CNV,
these values of $k$ correspond to 0.8\% or 3.2\% of the human genome
being CNV mutation hotspots, with CNV mutation rates similar to the
rates observed at hotspots of nonallelic homologous recombination \cite{20}.
We set an error rate for CNV typing of $\varepsilon= 0.75$. Based on
the resulting critical values, we calculated the power of observing a
significant result at $\alpha= 0.05$, assuming that \textit{de novo}
CNVs occur in $p = 1\%$, 0.5\% or 0.25\% of all cases at a particular
risk locus. These values are consistent with reports that \textit{de
novo} CNVs thought to contribute to the risk of psychiatric diseases are
observed in 0.2\% to 1\% of all cases \cite{20,60,61}. For comparison,
we calculated the total sample size necessary to achieve 80\% power in a
balanced case-control design at a significance level of $10^{-5}$ for
all values of $p$, assuming that the CNV has full penetrance and is not
observed among controls.

Our results (Table \ref{tab2}) indicate that observing a large number of CNV
mutations at a single locus is unlikely. Under all considered scenarios,
observing 6 or more \textit{de novo} CNVs at one locus constitutes a
significant result. For sample sizes $>$1000 trios, we have reasonable
power to detect \textit{de novo} CNVs that are present in 0.5\% of the
cases. The results indicate that the power of this approach strongly
depends on the total number of CNV mutation hotspots, which is unknown
for the human genome. However, even if only 500 CNV hotspots exist
genome-wide, for a CNV observed in 1\% (0.5\%) of all cases, only 1000
(2000) trios are necessary to achieve $\sim$80\% power. For comparison,
5500 (11,000) unrelated individuals are required to achieve similar
power. This suggests that testing for an excess of \textit{de novo}
mutations is a more powerful strategy than case-control testing.

\subsection*{CNV Load of Rare Variants}

As discussed previously, most minor alleles of CNVs are rare, and tests
of association between rare variants and a phenotype have limited power.
However, it is conceivable that multiple, independent CNVs each
contribute to disease risk. Jointly testing all CNVs may therefore be
more powerful than testing markers individually. If all such risk CNVs
cover the same genomic interval, the contribution of that region can be
determined by counting the number of individuals who carry a minor
allele of any CNV overlapping with the region. The counts in cases and
controls can then be compared \cite{62}, essentially treating all overlapping
CNVs as a single risk allele. Such joint analysis of multiple CNVs is
more challenging under a model of genetic heterogeneity, which assumes
that a large number of unlinked loci in the genome are affected by CNVs
that contribute to the risk of disease. Under this model, testing each
individual locus may not result in a significant signal. However, as the
total number of CNVs in an apparently healthy individual is small \cite{3},
if several CNVs contribute to the risk of disease, cases as a whole may
carry substantially more minor CNV alleles than do controls. Therefore,
a commonly applied test is to count the number of minor CNV alleles
observed genome-wide in cases, and compare that to the number of minor
CNV alleles in controls. While such a test is more powerful under a
model of genetic heterogeneity, it has two weaknesses. First, it is very
sensitive to any experimental error affecting cases and controls
differentially. For example, batch effects can increase the total number
of CNVs observed in one batch over the next. If cases and controls are
analyzed in different batches, such effects will immediately cause
significant genome-wide differences in CNV numbers between cases and
controls. Second, this test lacks interpretability. A significant signal
only indicates that CNVs somewhere in the genome increase disease risk.
Further testing of larger samples is required to understand the
contribution of individual CNVs. CNVs that show an excess of minor
alleles in cases that is not statistically significant may be good
candidates for further tests in larger samples.

This strategy has been used successfully to identify CNVs affecting
autism. Initially, Sebat et al. \cite{33} observed a genome-wide excess of
CNVs in autism patients, but no individual CNV or locus was significant
after multiple test correction. Further work by Weiss et al. \cite{61} in
larger samples demonstrated that several of the loci showing suggestive
evidence in the original report \cite{33} are significantly associated with
the risk of autism.

\section*{Conclusions}

Copy number variation accounts for much of the genetic variation
discovered to date in humans. Some of this variation is clearly
functional; studies in recent years have discovered several CNVs
contributing to the risk of common complex diseases such as autism~\cite{33}
and psoriasis \cite{31}. Hence, studying the contribution of CNVs to common
diseases has become standard practice during the course of genome-wide
association studies. As the genotype and hybridization signals generated
by genotyping platforms provide information about CNVs, it is efficient
to use this signal to infer CNV location and carrier status in the
sample. Most studies that impute CNVs from genotype array data focus on
analyzing hybridization data. Such analyses require careful
normalization of the intensity data, as hybridization signals are
susceptible to experimental noise that may lead to false inferences.

Many methods have been developed to localize CNVs using such
hybridization intensity data (e.g., \cite{15,48}). While such methods
generally perform well for CNVs covered by large numbers of probes, they
tend to have higher error rates for CNVs that span only a few genotyped
probes \cite{15}. The high error rates are in part due to the fact that these
methods aim to jointly localize CNVs and determine carrier status in
individuals, thus increasing the uncertainty of the inference procedure.
As databases now contain $>$20,000 CNVs, many of which have been observed
more than once in different samples, it seems likely that most common
CNVs are now known and that CNVs not yet present in databases are rare,
at least in Caucasians. Consequently, algorithms have been developed
that, using GWAS data, infer copy number for CNVs whose boundaries are
known \cite{38,43}. Such algorithms are more precise in calling common CNVs
and therefore facilitate more powerful tests for association. However,
it is still necessary to apply more general algorithms to detect unknown
CNVs, rare CNVs and \textit{de novo} events.

Most CNVs are in strong LD with SNPs in the HapMap, and it is not always
clear that inferring CNVs to test for association is the most powerful
strategy. As we have shown, even modest error rates in CNV calling
result in a loss of power comparable to testing a tag SNP with $r^{2}
\leq0.8$. Consequently, tests based on inferred CNV alleles are often
unlikely to be more powerful than testing surrounding SNPs for
association. In this context, it is interesting to note that both common
risk CNVs identified to date through genotype scans were first localized
via tag SNPs; only follow-up testing identified these CNVs as likely
risk alleles \cite{30,32}.

On the other hand, this observation also indicates that SNPs flanking a
CNV can provide information about CNV carrier status, suggesting that
existing methods for calling CNV alleles could be improved by jointly
considering haplotype background and hybridization intensity of the
covered markers. As haplotype background and hybridization intensity
provide orthogonal evidence for CNV status, such a method would likely
be substantially more precise and allow more powerful tests.

Maximizing the power of tests for association is crucial, as the effect
sizes of common CNVs are likely to be small, and the minor alleles of
most CNVs are rare. Hence, tests for association between CNVs and
diseases are likely to have low power even under the best circumstances.
Other strategies to identify CNV contributing to the risk of common
diseases should also be explored. Recent studies have indicated that
\textit{de novo} events generating new CNVs are rare \cite{3}. This suggests
that testing a genomic region for an excess of \textit{de novo} CNVs is
potentially a powerful strategy. Moreover, such \textit{de novo} CNVs
are more likely to have large effect sizes. Several such CNV regions
have in fact been detected \cite{33,34}. We have presented calculations,
based on conservative estimates, that for moderate sample sizes, simple
tests for local excesses of \textit{de novo} mutations have good power
to identify such CNV mutation hotspots. However, commonly collected
samples of unrelated cases and controls do not provide any information
as to whether a CNV observed several times among cases and not in
controls is the result of several \textit{de novo} mutations or just a
result of sampling variation on a rare CNV. Hence, the power to detect a
rare risk variant that is the result of multiple \textit{de novo} events
is substantially higher in family-based studies.

Any CNV-disease association identified using genotyping chips has to be
evaluated carefully. While analyzing the hybridization signal from
genotyping platforms provides cheap information about CNV status in a
population, genotyping arrays are not the gold standard for determining
carrier status. In association studies, erroneous calls of carrier
status usually result in a loss of power, rather than false positive
associations. However, in a study testing for an excess of \textit{de
novo} mutations, even a few inaccurate CNV calls can lead to false
positives. Hence, if a CNV appears to be associated with a phenotype, it
seems prudent to use an independent technology, such as CGH, PCR or
resequencing, to verify the inferred carrier status and the association
signal.

In the near future, GWAS will be supplemented with studies that sequence
regions of interest or even the entire genomes of affected and
unaffected individuals. Paired end-sequencing allows the detection of
regions where the distance between two short reads is significantly
longer or shorter than expected; such regions are likely to carry CNVs.
In the case of single-end sequencing, it is necessary to infer CNVs by
indirect measures, such as the number of reads generated for each
region. As read-lengths increase, it becomes easier to align reads, and
it may become possible to identify CNV breakpoints within a read. Then
it will be possible to identify such features from the generated
sequence. As technology advances at an incredible pace, we will
constantly be challenged to develop newer, better statistical tools to
infer the presence and location of CNVs.

Ultimately, to understand the role that CNV plays in human disease, we
must better elucidate the biological processes that create CNV, improve
the sensitivity and specificity of experimental methods that identify
CNV, and develop statistical methods that fully leverage the signals of
CNV that exist in data derived from genome-wide genotyping arrays as
well as next-generation sequencing technologies.

\begin{appendix}\label{app}
\section*{Appendix}

In the following we will explore patterns of genotype frequencies and
the Hardy--Weinberg equilibrium in SNPs covered by CNVs. We consider the
observed haplotype frequencies $P(G=AA)$, $P(G=AB)$, $P(G=BB)$ of a SNP with
alleles $A$ and $B$. Let $p$ be the population frequency of allele $A$ and $q$
the allele frequency of allele $B$. We first consider deletions, then
duplications.

\subsection*{Deletions} We introduce the segregating deletion as
a third genotype $D$ with frequency $d$, so that $d+p+q=1$. Let
$G\in\{AA,BB,AB\}$ indicate the possible observed genotypes and
$T\in\{DD,AA,AD,AB,\break BD,BB\}$ be the set of all possible true genotypes.
Genotype $DD$ will result in a failed genotyping reaction and hence will
never be observed.

Then, assuming no genotyping error for nondeletion alleles,
\begin{eqnarray*}
P(G=AA)&=&P(T=AA|T\neq DD)\\
&&{}+P(T=AD|T\neq DD)\\
&=&(p^{2}+2 dp)/(1-d^{2}),
\\
P(G=BB)&=&P(T=BB|T \neq DD)\\
&&{}+P(T=BD|T\neq DD)\\
&=&(q^{2}+2 dq)/(1-d^{2}),
\\
P(G=AB)&=&P(T=AB|T\neq DD)\\
&=&2pq/(1-d^{2}).
\end{eqnarray*}
Note that the estimated frequency of allele $A$ is
\begin{eqnarray*}
p_{\mathrm{est}}&=&P(G=AA)+1/2P(G=AB)\\
&=&(p+dp)/(1-d^{2})=p/(1-d),
\end{eqnarray*}
so that we will observe fewer than the expected number of heterozygotes,
given the estimated allele frequencies:
$2p_{\mathrm{est}}q_{\mathrm{est}}=2pq/(1-d)^{2}\geq2pq/(1-d^{2})=\mathrm{E}(G=AB)$.

In a sample of n individuals with observed genotype counts ($C_{AA}$,
$C_{AB}$, $C_{BB}$), the expected departure from HWE is
\begin{eqnarray*}
&&\mathrm{E}\bigl((C_{AB})^{2}-4C_{AA}C_{BB}\bigr)\\
&&\quad=\operatorname{Var}(C_{AB})+\mathrm{E}(C_
{AB})^{2}\\
&&\qquad{}-4[\mathrm{E}(C_{AA})\mathrm{E}(C_{BB})+\operatorname{Cov}(C_{AA},C_{BB})]
\\
&&\quad=npq/(1-d ^{2})+8 ndpq(1-n)/(1-d).
\end{eqnarray*}

\subsection*{Insertions} We consider a model where every haploid
copy of the genome carries 0 or 1 insertions. Let $T\in\{A,B\}$ indicate
the true genotype at the original location and $I\in\{A,B\}$ indicate
the genotype at the duplication. To account for possible LD between the
inserted region and the original copy, as insertions can happen multiple
times, we consider the probability of carrying the insertion allele
conditional on the allele at the original location, $Q(I|T)$. We define
$D\in\{0,1,2\}$, the total number of insertions on both chromosomes, and
the probability of carrying a duplication is $d_{T}$:
\begin{eqnarray*}
&&P(G=AA)\\
&&\quad=P(T=AA)[P(D=0|A)\\
&&\qquad\hphantom{P(T=AA)[}{}+P(I=A)P(D=1|A)\\
&&\qquad\hphantom{P(T=AA)[}{}+P(I=AA)P(D=2|A)]
\\
&&\quad =p^{2}[(1-d_{A})^{2}+2 d_{A}(1-d_{A})Q(A|A)\\
&&\qquad\hspace*{93pt}{}+d_{A}^{2}Q(A|A)^{2}]\\
&&\quad=p^{2}\bigl(1-d_{A}Q(B|A)\bigr)^{2},
\\
&&P(G=BB)=q^{2}\bigl(1-d_{B}Q(A|B)\bigr)^{2}.
\end{eqnarray*}
Hence, $P(G=AA)=P(T=AA)=p^{2}$ iff $Q(A|\break A)=1$ or $d_{A}=0$. Similarly
$P(G=BB)=P(T=BB)=q^{2}$ iff $Q(B|B)=1$ or $d_{B}=0$. No departure from
HWE is generated if all duplications with the~$A$ allele occur on
chromosomes carrying the $A$ allele at the original location (likewise for
allele $B$), regardless of the frequency of the insertion. This pattern
may be observed if duplications are the result of nonallelic homologous
recombination or other mechanisms creating tandem repeats.
\end{appendix}
\section*{Acknowledgments}

We would like to thank Margit Burmeister, Don Conrad and Jun Li for
thoughtful readings of the manuscript. SZ is supported by HL084729-02;
TMT is supported by DK062370 and T32 HG00040.

\end{document}